-RESEARCH ARTICLE-

# Online LDA based brain-computer interface system to aid disabled people

Apdullah Yayık[1*] and Yakup Kutlu[2]

[1] Institute of Alparslan Defence Sciences, National Defence University, Ankara, Turkey
[2] Department of Computer Engineering, İskenderun Technical University, Hatay, Turkey

**Abstract**
This paper aims to develop brain-computer interface system based on electroencephalography that can aid disabled people in daily life. The system relies on one of the most effective event-related potential wave, P300, which can be elicited by oddball paradigm. Developed application has a basic interaction tool that enables disabled people to convey their needs to other people selecting related objects. These objects pseudo-randomly flash in a visual interface on computer screen. The user must focus on related object to convey desired needs. The system can convey desired needs correctly by detecting P300 wave in acquired 14-channel EEG signal and classifying using linear discriminant analysis classifier just in 15 seconds. Experiments have been carried out on 19 volunteers to validate developed BCI system. As a result, accuracy rate of 90.83% is achieved in online performance.

**Keywords:**
Brain-computer interface, P3 wave, Event-related potentials

## Introduction

For severe disabled people, assistive technology is very significant aid. It not only let them to make their life more independent but also increases quality of their life. People who suffer from spinal cord injury (SCI), amyotrophic lateral sclerosis (ALS) or guillain-barre syndrome needs assistive technology to make their life survived. Most of these have not ability to perform daily activities without any help. Global communities are trying to help disabled people meet with their needs. Also, researchers are currently investigating adaptive technologies that can aid them particularly in controlling computer and accessing information (Subaşıcıoğlu, 2008). Researchers have been investigating different approaches to develop effective assistive technologies since 2 decades. The principal part of these approaches is to be able to convey commands and communicate with other people without any motor muscular system.

---

* *Corresponding Author, email: apdullahyayik@gmail.com*

Several ways are used by researchers for interacting without motor muscular movements. Electrooculography (EOG) based human-computer interfaces (HCI) that are able to determine the eye movement direction in 2 dimensions. This approach is used to control computer (Borghetti, Bruni, Fabbrini, Murri, & Sartucci, 2007) and to advance mouse cursor control system (Surdilovic & Zhang, 2006).

Another promising approach is brain-computer interface (BCI) systems that rely on the use of signal acquired from brain. BCI systems are field of adaptive technology that make them feasible for both disabled and able-bodied people not only to interact with each other but to control their daily activations such as navigating computer cursor, gearing up/down in a car, guiding drone and more. Unlike conventional control systems (speaking, writing, touching, seeing and more) that necessitates motor ability, BCI systems are controlled by "Brain" that serves as the centre of the nervous system and do not necessitates any motor ability. BCI systems include not only either neurology or computer science also physiology, biology and mathematics, because of complexity of this interdisciplinary research area. Measuring and exploring brain activity and dynamics include invasive and non-invasive amplifying techniques. While invasive techniques amplify signal from deeply placed electrodes on the brain with high cost and surgical intervention, non-invasive techniques amplify signal from scalp surface with low cost and non-surgical intervention. BCI systems prefer non-invasive brain signal Electroencephalography (EEG) due to its low setup cost, good temporal resolution, simplicity of use and mobility. In 1924 Berger amplified EEG signal from human's scalp which was the leading attempt of researches based EEG (Masson & Berger, 1924)..

In 1964 Walter averaged the EEG signal and revealed the brain response called event-related potential (ERP). He showed the first ERP component named contingent negative variation (CNV). This is the pioneering study that investigates ERPs (Walter, Cooper, Aldridge, & McCallum, 1964). ERPs are electrophysiological response that is voltage fluctuations to auditory, visual or tactile stimulus (Mesulam & others, 2000). ERP is a wave form that can be amplified from the surface of human scalp and revealed from online EEG after filtering and signal averaging processes. ERPs extracted from EEG provide evidences about information cognitive processing in human brain that is needed in performing BCI systems. Mainly there are two types of ERPs; exogenous and endogenous components. While exogenous component is a compulsory response of brain evoked by physical (light, sound and more)activity in external world, endogenous component is a processing response of brain like mental activities that is/isn't evoked by external world (Handy, 2005).

In 1965 Sutton et. al detected that when subject has no prior knowledge about stimuli (light or sound), latency of reaction in EEG signal is longer than when type and dynamics of stimuli is known before. This is the pioneering study that investigates P300 (P3) wave and oddball paradigm (Sutton, Braten, Zubin, & John, 1965). P3 wave is an ERP endogenous component elicited during decision making process. The most common and validated method is implementing the oddball paradigm, in which low-probability target items are mixed with high-probability non-target items (Berlad & Pratt, 1995). P3 wave is a positive deflection in voltage and can be evoked approximately 250 to 500 ms after the stimulus onset (Sutton et al., 1965). While latency of P3 wave elicited by visual stimuli is 350 to 450 ms, latency of P3 wave elicited by auditory or tactile stimuli is 150-300 ms. P3 wave is widely used in BCI systems; mouse control (Citi, Poli, Cinel, & Sepulveda, 2008), spelling (Rakotomamonjy & Guigue, 2008; Rivet, Souloumiac, Gibert, & Attina, 2008; Ron-angevin & Silva-sauer, 2006), robotic arm control (Potentials, 2011), visual object detection (Hoffmann, Vesin, Ebrahimi, & Diserens, 2007; Kutlu, Yayik, Yildirim, & Yildirim, 2015). In 2010 Donnerer M. and Steed A. investigated using P3 wave from 8 channel EEG signals in 3-dimentional (3D) environment and proved its possibility by developing a BCI system in 3D (Donnerer & Steed, 2010). Objects paradigm, tiles paradigm and 36 spheres paradigm were tried to integrate BCI to an augmented reality style control system. 36 spheres is exactly the same paradigm that is known as single character flashing in conventional BCI $6 \times 6$ sized speller scenario. In 2010 Townsend et. al proposed checkerboard paradigm to enhance BCI based $8 \times 9$ sized speller system (Townsend et al., 2010) that relies on P3 wave from 32-channel 256 Hz. EEG signals. The system was tested on 18 subjects and compared with row/column paradigm introduced by Farwell and Dolchin in 1988 (Farwell & Donchin, 1988). As a classifier stepwise LDA was preferred. As a result, it was seen that proposed checkerboard paradigm is superior to row/column paradigm. In 2010 Su et. al introduced BCI based Chinese speller system that uses P3 wave form 14-channel 250 Hz. EEG signals. The system was tested on 3 subjects. The important property of this study is channel selection is implemented with three different methods, rough set, F-score and accuracy of Fisher's LDA classifier both for Fisher's LDA and SVM classifiers. Rough set is selected as best one. As a result, channels that have high impact on classifiers were detected especially in midline and posterior of brain (Su et al., 2010). In 2011 Postelnicu et. al focused on controlling robotic arm with XVR (eXtreme Virtual Reality) software using 256 Hz. only 1 channel (O2) EEG and electrooculography (EOG) signals that could aid disabled people. The system tested on 8 subjects could convey 4 type of commands; left, right, top and bottom (Postelnicu, Talaba, & Toma, 2011). In 2011 Ashari et al. developed BCI based virtual telephone keypad that relies on SSVEP (Ashari, Al-Bidewi, & Kamel, 2011). In 2012 Kaufmann et. al investigated whether N2 wave of ERP can be used in a BCI based $6 \times 6$ speller system, or not. They reached that the best way was using only P3 wave or both P3 and N2 waves, not just N2 wave. 12-channel 256 Hz. EEG signals were used and the given results were proven with tests on 51 subjects. Important properties of this study are, flash duration (31.25 ms) and inter stimulus duration (125 ms) are very short and filtering range (0.1 Hz. and 30 Hz.) of raw EEG signals is broad. For denoising raw EEG signals any operation was performed except for filtering. It is seen that though high overlapping on ERPs and not-denoising raw EEG signals, successful results were reached (Kaufmann & Hammer, n.d.). In 2012 for a novel user interaction, Esfahani and Sundararajan achieved to distinguish primitive shapes (cube, cylinder, pyramid, sphere or cone) using 14-channel mental imagery EEG signals with features extracted by independent component analysis (ICA) and hilbert transform and linear discriminant analysis (LDA) classifier (Esfahani & Sundararajan, 2012). In 2012 Jin et. al proposed BCI based speller system that relies on both P3 wave and motion-onset visual potential. In this combined system visual stimuli effects includes colour change (blue/green, white/grey), moving (with $6 \text{ m/s}$ speed in 3 cm distance) and both. The system was tested on 10 subjects. EEG signals were recorded from 12 channels and sampled at 256 Hz. The researchers preferred downsampling EEG signals to 36.6 Hz. for extracting meaningful features, and bayesian LDA classifier. Also three different range of durations (0-800 ms, 0-299 ms. and 300-800 ms.) were tried to segment EEG signals. As a result, in online and offline experiments practical feedback was observed (Jin, Allison, Wang, & Neuper, 2012). In 2012, Long et. al investigated hybrid using of P3 wave and motor imagery to control computer cursor that performs moving on the screen and selecting/rejecting target with 30-channel EEG signals. They showed that hybrid approach was more effective than both P3 wave and motor imagery (Long, Li, Yu, & Gu, 2012). In 2012, Blasko et. al investigated developing three different visual stimuli based BCI systems for different purposes, using only P3 wave and only N2PC (N2-posterior-controlateral) wave and both of P3 and N2PC waves of EEG signals from 16-channels . These were exploring on internet browser, controlling robotic arm and a commination tool. As a result, it was reached that accuracy of both types of waves was relatively higher than that of others (Blasco, Iáñez, Ubeda, & Azor'\in, 2012). In 2012 Bougrain et. al investigated controlling a JACO robotic arm using motor imagery based BCI that uses 512 Hz. EEG signals from 13 channels and LDA classifier (Bougrain, Duvinage, & Klein, 2012). Four different motor imageries (left hand, right hand, both hands and feet) were used to convey four commands (left move, right move, forward, backward). The important property of this study is inverse reinforcement learning that focuses on post-processing of BCI system (Bougrain et al., 2012). In 2012 Shi et. al introduced a novel paradigm BCI based speller system, sub-matrix based paradigm. In this study, $6\times6$ sized matrix was divided into 6 number of $2\times3$ sub-matrices and $6\times9$ sized matrix was divided into 6 number of $3\times3$ sub-matrices and their performances were compared. The system was tested on 7 subjects with SVM classifier using 4,6,8 and 10 electrode configurations and it was seen that sub-matrix based paradigm was superior than row/column based paradigm, expansion of size in sub-matrix based paradigm did not make the system slower and 10 electrode configuration gives better results than others (Shi, Shen, Ji, & Du, 2012). In 2013, Velasco-Álvarez et. al investigated the use of wheel-chair on 6 subjects via EEG signals from 16-channles using navigation paradigm (that uses motor imaginary), in particular, without visual -stimuli (Velasco-Álvarez, Ron-Angevin, da Silva-Sauer, & Sancha-Ros, 2013). As a result, they showed online successful implementation. In 2013 Kaplan et. al surveyed exiting BCI games (chess, mindgame, bacteria hunt, brain invaders, mind the sheep! and a face card game) relies on P3 wave and explained how shortcomings were handled (Kaplan, Shishkin, Ganin, Basyul, & Zhigalov, 2013). In 2014 İşcan and Dokur introduced novel BCI system that is character plotter. It is really an advanced version of conventional BCI speller system. 6 circles were used to draw character, and 3 circles were used for drawing procedure (undo, raise and ok). The system was tested on 16 subjects. Steady-state visual ERP of 500 Hz EEG signals from 16-channels are used with SVM, LDA and nearest mean classifiers. The novelty of this study is, innovative speller, named as character plotter (İcscan & Dokur, 2014). In 2014 Vourvopoulos and Liarokapis investigated usefulness and effectiveness of commercial EEG amplifier devices in a robot navigation BCI application. They reached that Emotiv (tested on 31 subjects) and Neurosky (tested on 54 subjects) headsets are useful and effective (Vourvopoulos & Liarokapis, 2014). In 2014 Tsuda et. al presented BCI based visual object detection using P3 wave of 4-channal 256 Hz. EEG signals and oddball paradigm. As visual stimulus 4 different images are used. LDA, k-NN and nearest mean classifier are applied and compared (Tsuda, Lang, & Wu, 2014). In 2014 Koo et. al proposed a combined human computer interface (HCI) system that relies on both P3 wave of EEG and EOG signals. In this system two monitors exist and monitor selection is performed by EOG signals and task selection is performed by P3 wave of EEG signals. Originality of this study is requiring half time for decision because of being a combined system. 8-channel 2048 Hz. sampled EEG signals were used. The system was successfully test on only 1 subject with SVM classifier (Koo, Nam, & Choi, 2014). In 2015 Bai et. al presented a hybrid BCI system, relies on both P3 wave and motor imagery of 250 Hz. EEG signals, could operate an explorer. The system included BCI mouse, BCI speller and an explorer. The system was tested on 5 subjects with SVM classifier and promising results were reached (Bai, Yu, & Li, 2015). In 2015 Combaz and Hulle proposed combining P3 and steady-state visual evoked potential (SSVEP) to receive better performance than when each one is used purely. To see results, three different experiments (pure P3, pure SSVEP and combined) were designed. While flashings colour of target changes grey to yellow. 32-channel 1024 Hz. EEG signals were classified using SVM classifier. The system was tested on 25 subjects (9, 8 and 8 of them participated in experiment 1, 2 and 3, respectively). As a result it was reached that performance of combined potential was relatively higher than that of others. Important properties of this study are providing synchronizing between display and recording times, Psychophysics Toolbox (Brainard, 1997) Matlab extension was used and selecting train and test sets using leave-one-out cross validation (Combaz & Van Hulle, 2015).

Main objective of this paper is to construct a non-invasive BCI system based on P3 wave that enables neurologically disabled people to convey their needs and to communicate with other people easily and effectively. The system is based on one of the most effective event-related potential (ERP) wave, P3, which can be elicited by oddball paradigm. Developed application has a basic interaction tool that enables disabled people to convey their needs to other people selecting related objects. These objects pseudo-randomly flash in a visual interface on computer screen. The user must focus on related object to convey desired needs. The system can convey desired needs correctly by detecting P3 wave in acquired 14-channel EEG signal and classifying using linear discriminant analysis (LDA) classifier just in 15 seconds. Experiments have been carried out on 19 volunteers to validate developed BCI system. An accuracy rate of 90.83% is achieved in online performance.

The remaining of this paper is organized as follows; in Section 2 experimental setup and database information are performed. Pre-processing and classification algorithms are described in Section 3. The experimental results obtained with 19 volunteers are given in Section 4. Conclusion and discussions of results are performed in Section 5.

**Material and methods**

*Database Description*

19 no-paid male volunteers (aged between 20-32) participated in this study. All of the subjects did not have any BCI application experiences. In the course of EEG recordings, subjects were asked to relax and refrain movements that are redundant.

EEG signals were amplified with Emotiv EPOCH+ with 128 Hz sampling frequency. EEG signal was recorded from 14 channels from standard locations of 10-20 international system. EEG electrodes are AF3, F7, F3, FC5, T7, P7, O1, O2, P8, T8, FC6, F4, F8, and AF4.

*Experimental Setup*

Experimental scenario is designed using OpenVIBE designer software (Renard et al., 2010) and to provide parallel processing between scenario and EEG amplifier, OpenVIBE acquisition server is used in a different computer as seen in Figure 2. At the beginning of experiment, each subject were asked to fill poll to learn about personal knowledge, any neurologic disease experience and cigarette, alcohol or any drug habits. The experiment didn't begin for the subject who has neurologic disease or is drug habit because of negative effect on P3 wave's both latency and amplitude. Subjects were seated about 100 cm in front of $30 \times 48$ cm sized LCD screen on which twelve images related to scenario that were on $4 \times 3$ matrix in Figure 1 are displayed. The images shows; house, television, telephone, car, bed coffee, meal, bath, shopping cart, internet modem, popcorn and heart. The images selected for system in which subjects will be able to request help using developed BCI. In P3 paradigm each flash lasts 200 ms ($d_{flash}$) and in course of next 100 ms ($d_{non-flash}$) no images flash.

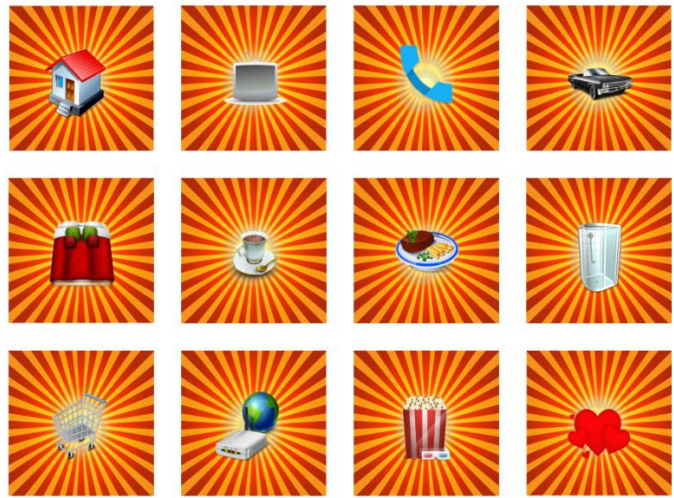

**Figure 1.** Images shown in the experimental scenario.

The images flashes block-randomly to avoid double flashings, one by one (Fazel-Rezai, 2007). So interstimulus interval is 300 ms. EEG laboratory layout is shown in Figure 2. Signal processing and machine learning algorithms are implemented in MATLAB EEGLab toolbox (Delorme & Makeig, 2004) and WEKA data mining tool (Hall et al., 2009).

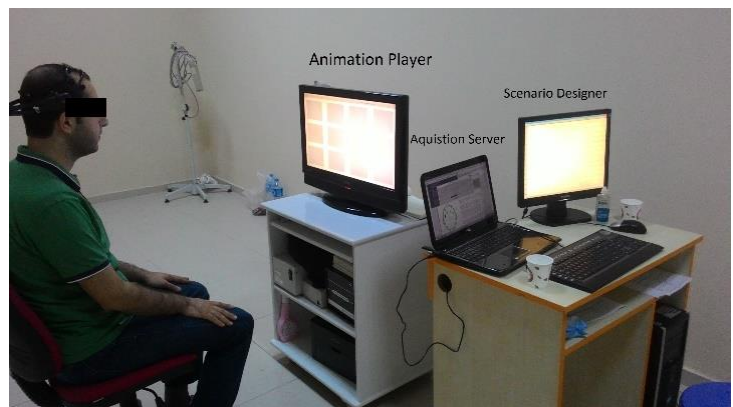

**Figure 2.** EEG laboratory layout.

*Experimental Schedule*

All the participants had been informed about EEG recording procedure and ethical issues in the Declaration of Helsinki, World Medical Association (VMA)-Ethical Principles for Medical Research Involving Human Subjects (Association & others, 2001), before experiment started.

- In this study, each subject completed an experiment that includes 6 scenarios. And, scenario includes 12 sessions each of which has 6 runs. Following schedule is used while recording.
- When experiment started in adaptation duration of 10 s ($d_{adapt}$) all of the images are shown and amplified EEG are not recorded.

- Image that subject will follow during each run is shown as information in 3 s ($d_{inf}$). Aftermath of each run subjects whispered counted number in previous run.
- Subject counts how often prescribed image flashes in each run. If subject didn't count accurately, recording removed and experiment was repeated after warning for attention.
- Random sequence of flashing starts and EEG is being recorded simultaneously.
- Interval of runs is 0.2 s ($d_{run-interval}$).

Duration of each run is,

$$d_{run} = (d_{flash} + d_{no-flash}) \times 12 \quad (1)$$
$$= (0.2 + 0.1) \times 12$$
$$= 3.6\,s$$

Duration of each session completed for each image is,

$$d_{session} = d_{inf} + (6 \times d_{run}) + (5 \times d_{run-interval}) \quad (2)$$
$$= 3 + (6 \times 3.6) + (5 \times 0.2)$$
$$= 25.6\,s$$

Duration of scenario is,

$$d_{sce} = d_{adapt} + 12 \times d_{session} \quad (3)$$
$$= 10 + 12 \times 25.6$$
$$= 317.6\,s$$
$$\cong 5\,\min 18\,s$$

**Pre-processings**

**Filtering**

Although P3 wave particularly exists between 0.1 Hz and 4 Hz (Jin et al., 2012), due to possibility of its existing in higher frequencies (Kolev, Demiralp, Yordanova, Ademoglu, & Isoglu-Alkaç, 1997) cut-off frequencies is preferred as higher than 4 Hz. 3rd order butterworth (Butterworth, 1930) band pass filter that has 0.1 Hz low cut-off frequency and 20 Hz high cut-off frequency is designed to filter data.

**Feature Scaling**

In this thesis, to suppress the effect of outliers, in all classifiers feature data is scaled to 0 and +1 range using min-max normalization method before classification step.

*Segmentation*

For each subject;

- Each run has 12 segments,
- Each session has $12 \times 6 = 72$ segments,
- Scenario has $12 \times 6 \times 12 = 864$ (72 of which had P3 wave and 792 of which has not P3 wave.)

segments.

**Feature Vector Construction**

Data acquired from channel FC5 is removed because of having very high number of NaN values. Therefore, remaining process is implemented for 13 EEG channels. The segments obtained from EEG channels are concatenated and used as features. Features size is

$$\text{Feature Size} = 1 \times (D_c \times D_s) \tag{4}$$

where $D_c$ is number of channels (13) and $D_s$ number of EEG data in segment.

$$D_s = (\text{Sampling Frequency}) \times (\text{Duration of Single Trial}) \tag{5}$$
$$= 128 \times 0.5$$
$$= 64 \text{ (for avaraging it is preferred as 65)}$$

So, each feature vector size is $13 \times 65 = 1 \times 845$

**Independent Component Analysis (ICA)**

Independent Component Analysis (ICA) is implemented to disassemble the band passed filtered EEG signal into artefacts and spontaneous EEG activity. Therefore STNR can be enhanced and clearer signal can be revealed (Makeig, Bell, Jung, Sejnowski, & others, 1996). ICA is an algorithm group which tries to achieve blind separation of sources. Observed EEG signal is,

$$x(t) = As(t) \tag{6}$$

where $s(t)$ is statistically independent sources which is unknown and A is mixing matrix that is unknown but invertible. The aim of ICA algorithm is to calculate $u(t)$ that is recovered version of $s(t)$.

$$u(t) = Wx \tag{7}$$

where $W$ is de-mixing matrix that is equal to $A^{-1}$ and allows to calculate recovered version of source signal. In this thesis, FastICA algorithm (Hyvärinen, Karhunen, & Oja, 2004) is used for ICA implementation.

*Linear Discriminant Analysis (LDA)*

The classifier is responsible for deciding on whether acquired signal is produced while focusing on target or not. In this paper linear discriminant analysis (LDA) classifier is used for this process. LDA is mostly used for separating two classes in a multidimensional space. This method is not computationally intensive because it is based on the 1$^{st}$ and 2$^{nd}$ moments of each distribution. In order to evaluate the minimum classification error the two classes are Gaussian with equal covariance (Mika, Ratsch, Weston, Scholkopf, & Mullers, 1999). It is a limitation of LDA classifier. The method is based on projection of high-dimensional data onto a line and classifies in a one dimensional space. Assume that the feature vectors be denoted by x. LDA is defined as the linear function $y = w^T x$ which maximizes the criterion function (Duda, Hart, & D.G., 2012). Sub-indices denote the groups.

$$J(w) = \frac{|\mu_1^2 + \mu_2^2|^2}{(S_1^2 + S_2^2)} \tag{8}$$

where $(S_1^2 + S_2^2)$ is the total within-class scatter of the projected samples and $|\mu_1^2 + \mu_2^2|$ is the difference of the projected means.

**Experimental Results**

Experiments with 19 volunteers have been carried out to validate efficiency and availability of BCI system. It is very important that before performing online experiment, with existing EEG data subject-based training process is performed for each volunteer for 2 times, as shown in Figure 3. First training is performed with acquired EEG data during scenario and classifier coefficients are saved. Then online BCI system is implemented using these coefficients and at the same time EEG data is acquired and saved.

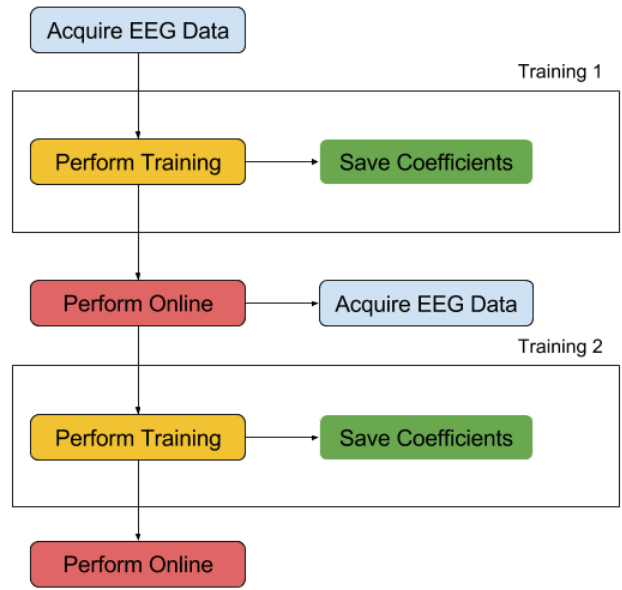

**Figure 3.** Performed training process for each volunteer.

At this point, low online performance is observed due to timing mismatch between acquiring and online scenario software. In order to handle this problem second training is performed with EEG data saved during online scenario and classifier coefficients are saved. Then with secondly calculated coefficients online scenario is performed and higher performance is achieved. It must be mentioned that during first online scenario flashing target sequence of objects are exactly same as in first acquiring process, while second online scenario can be fully independent.

Using constructed online scenario after 2 training process, at first volunteers self-decides the object related to desired command then focus on it and count how many times it flashes. LDA classifier classifies online acquired EEG signal as target or non-target using P3 wave in 3 trials. In each trial this classification process is performed, then classifier votes majority one as selection. As a response selected object appears in computer and relates sound file is played. For example when "car" object is selected, "Get my chauffeur prepare my car" sound plays.

120 objects, which includes 10 number of each objects in Figure 1, are given to volunteers and asked them to select each one in turn. 3 trials are implemented, so volunteers always counted 3 when focusing on desired object. And classifiers votes majority one in 3 decisions. Figure 4 shows mean number of correctly selected objects against each object in Figure 1, after BCI system is tested on each volunteer. 109 of 120 objects are correctly selected by volunteers, this mean developed BCI system has 90.83% accuracy rate in online performance.

For end user of this BCI system, training and saving coefficients takes 15 minutes and selecting any desired objects takes just 15 seconds.

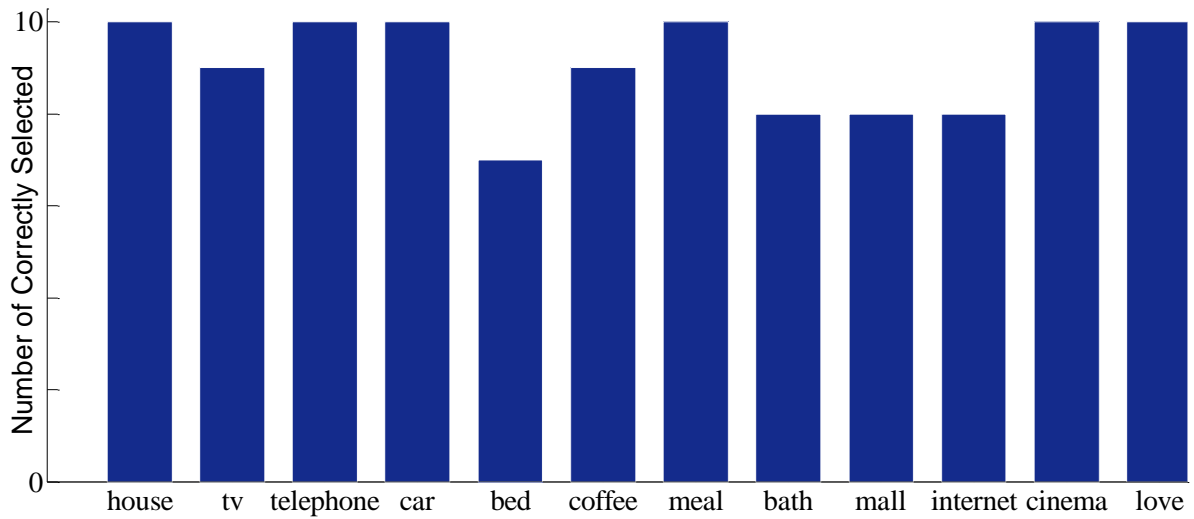

**Figure 4.** Number of correctly selected objects in online BCI system with 3 trials

**Discussion and Conclusion**

Objective of this paper is to advance a non-invasive BCI system based on P3 wave that lets neurologically disabled people to transmit their needs and to communicate with other people easily and effectively.

In Figure 4 objects that are in particularly at corners of screen (in Figure 1) are wrongly selected, while others are correctly selected by LDA classifier. It can be said that position of objects in visual screen highly affects online performance of BCI system. Accuracy rate of 90.83% can be increased via constructing different shapes and dimensions of visual objects.

The results show that developed BCI system is able to decide on the focused object in a very short time and conveys command easily. This work can inspire researches to advance robot control applications using proposed techniques.